\documentclass[prl,aps,twocolumn,showpacs]{revtex4}
\usepackage{graphics}% Include figure files
\usepackage{graphicx}% Include figure files
\usepackage{dcolumn}% Align table columns on decimal point
\usepackage{bm}% bold math
\usepackage{amsmath,amssymb}

\newcommand{\im}[1]{\textrm{Im}\left\{  #1\right\}}

\begin{document}
\title{Ultrashort self-induced transparency plasmon solitons}
\author{Andrea Marini$^1$ and Fabio Biancalana$^{1,2}$}
\email{andrea.marini@mpl.mpg.de}
\affiliation{$^1$Max Planck Institute for the Science of Light, Guenther-Scharowsky-Stra\ss e 1, 91058 Erlangen, Germany}
\affiliation{$^2$School of Engineering \& Physical Sciences, Heriot-Watt University, EH14 4AS Edinburgh, United Kingdom}
\date{\today}
\begin{abstract}
We study the interband self-induced transmission of surface plasmon polaritons in a gold film surrounded by an external
Kerr medium. We model the optical propagation by using a version of the generalized nonlinear Schr\"odinger equation for 
the field envelope coupled to Bloch equations for valence electrons of gold, predicting self-induced transparency of 
ultrashort plasmon solitons with a pulse duration below $10$ fs. We demonstrate that the Kerr nonlinearity from the 
surrounding dielectric can be used to compensate for the group velocity dispersion, and that the impact of dephasing and 
decay processes can be effectively reduced by the self-induced transmission mechanism. 
\end{abstract}
\maketitle

\paragraph{Introduction --} Surface Plasmon Polaritons (SPPs) -- exponentially localized electromagnetic waves guided along 
metallic surfaces -- are particularly important in a broad range of applications including medicine 
\cite{GobinNanoLett2007}, biosensing \cite{AnkerNatMat2008}, imaging \cite{KawataNatPhot2009} and nanophotonics 
\cite{GramotnevNatPhot2010}. A large variety of nonlinear mechanisms occurs in metals, e.g. second and third harmonic 
generation \cite{GinzburgNewJPhys2013}, nonlocal ponderomotive forces leading to symmetry breaking \cite{DavoyanOL2011}, 
and redshift due to interband thermo-modulation \cite{MariniNewJPhys2013}. The subwavelength confinement of SPPs is 
fundamentally important for designing nanoscaled optical circuits and for the enhancement of nonlinearity, which can in 
turn be exploited for active all-optical control \cite{KauranenNatPhot2012}, plasmon-soliton formation \cite{Feigenbaum2007} 
and nanofocusing \cite{DavoyanPRL2010}. Inevitably, these applications are either limited or even prevented by large 
intrinsic ohmic losses of metals. A possible strategy to retain the subwavelength localization of SPPs and overcome the 
loss barrier of metals at the same time is to embed gaining media in plasmonic devices \cite{GatherNature2010}. The 
intrinsic nonlinearity of two-level gaining media can then be used to obtain transverse localization of dissipative plasmon 
solitons \cite{MariniPRA2010}. Alternatively, losses can be limited by using long range surface plasmon polaritons 
(LRSPPs) in metallic films \cite{BeriniAdvOpt2009}, where localization is smaller and absorption is reduced accordingly. 
In common plasmonic setups operating at optical frequencies, it is likely to use silver as metallic component in order to 
minimize losses. Indeed, while the properties of silver can be practically understood from the free-electron plasma theory, 
the optical properties of gold and copper are greatly influenced by interband transitions, which boost absorption and damp 
optical propagation.  

\begin{figure}[b]
\centering
\begin{center}
\includegraphics[width=0.4\textwidth]{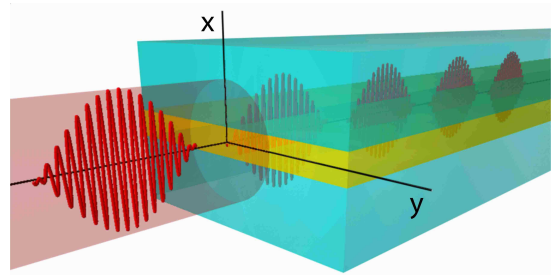}
\caption{Sketch of the representative structure analyzed in this work. Gold film of width $a = 20$ nm surrounded by an 
external Kerr medium. The structure is assumed infinitely extended in the $z,y$ directions.}
\label{StripeFig}
\end{center}
\end{figure}

In this Letter, we show the theoretical possibility to bypass interband absorption of LRSPPs in a thin film of gold 
surrounded by an external Kerr medium using self-induced transmitted $\pi$-pulses. Indeed, while {\it long} pulses 
($\tau \gtrsim 100$ fs) experience interband thermo-modulation \cite{MariniNewJPhys2013}, {\it ultrashort} 
pulses with time duration comparable with or smaller than the electron-electron collision time ($\tau \lesssim T_2 \simeq 10$ fs) 
do not undergo thermalization. Thus, ultrashort optical pulses can induce transient inversion of population and experience 
the effect of interband polarization. If one neglects non-conservative dephasing and recombination processes, perfect 
{\it self-induced transparency} (SIT) can be achieved with $\pi$-pulses \cite{McCallPhysRev1969}. 
Moreover, for some specific parameters, SIT pulses can coexist with temporal plasmon solitons, where the group velocity 
dispersion (GVD) is compensated for by the Kerr nonlinearity, analogously to what happens in erbium-doped fibers 
\cite{NakazawaPRL1991}. As a consequence of non-conservative processes, ideal transparency can not be achieved but 
enhanced self-induced transmission occurs, as we show here for the first time. A sketch of the setup considered in our 
calculations is depicted in Fig. \ref{StripeFig}.

\paragraph{Background --} The temporal dynamics of electrons in the presence of an external electromagnetic field is 
determined by the Schr\"{o}dinger equation with unperturbed Hamiltonian ${\cal H}_0$ and dipole interaction Hamiltonian 
${\cal H}_I(t)$. Since the electron momentum is conserved in the dipole interaction, it is possible to split the dielectric 
polarization of gold in {\it intraband} ($I$) and {\it interband} ($II$) contributions. For intraband transitions, due to 
the high occupation of states in the conduction band, the periodic potential of the lattice is screened by the quasi-free 
conduction electrons. The unperturbed Hamiltonian can then be approximated by the kinetic part 
${\cal H}_0 \approx p^2/(2m_{\rm e})$ and thus the dielectric polarization is ${\bf P}_I = \epsilon_0 \chi_I {\bf E}$, 
where $\epsilon_0$ is the dielectric permittivity of vacuum, 
$\chi_I = - \omega_{\rm p}^2/[\omega(\omega + \mathrm{i} \Gamma)]$, $\omega_{\rm p}=13.8\times 10^{3}$ THz is the plasma 
frequency of gold, $\Gamma =100$ THz is the damping rate due to electron-electron collisions, 
${\bf E} = {\bf E}_0 {\mathrm e}^{- \mathrm{i} \omega t}$ is the electric field and $\omega$ is its angular frequency.  

\begin{figure}
\centering
\begin{center}
\includegraphics[width=0.2\textwidth]{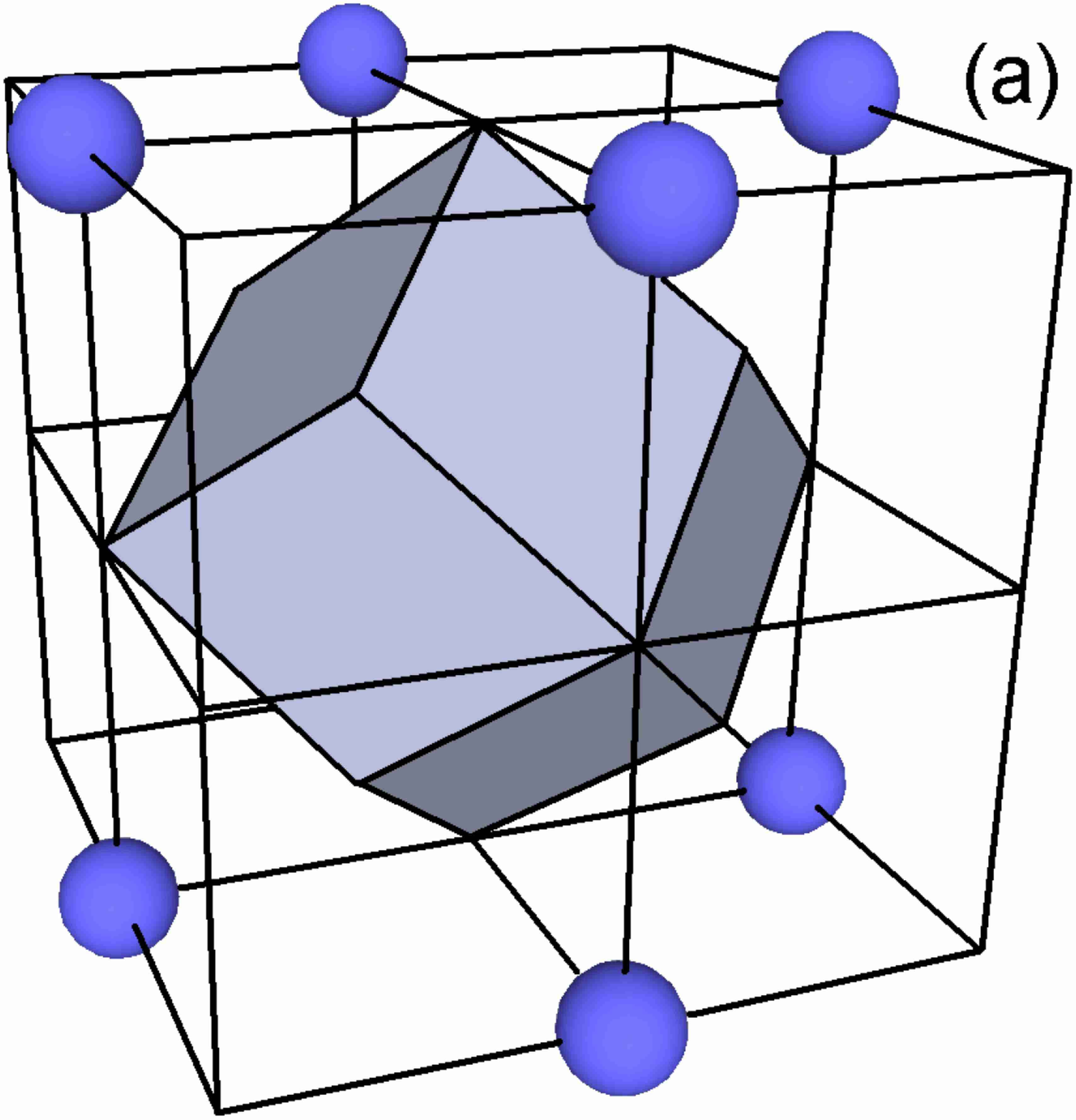}
\includegraphics[width=0.2\textwidth]{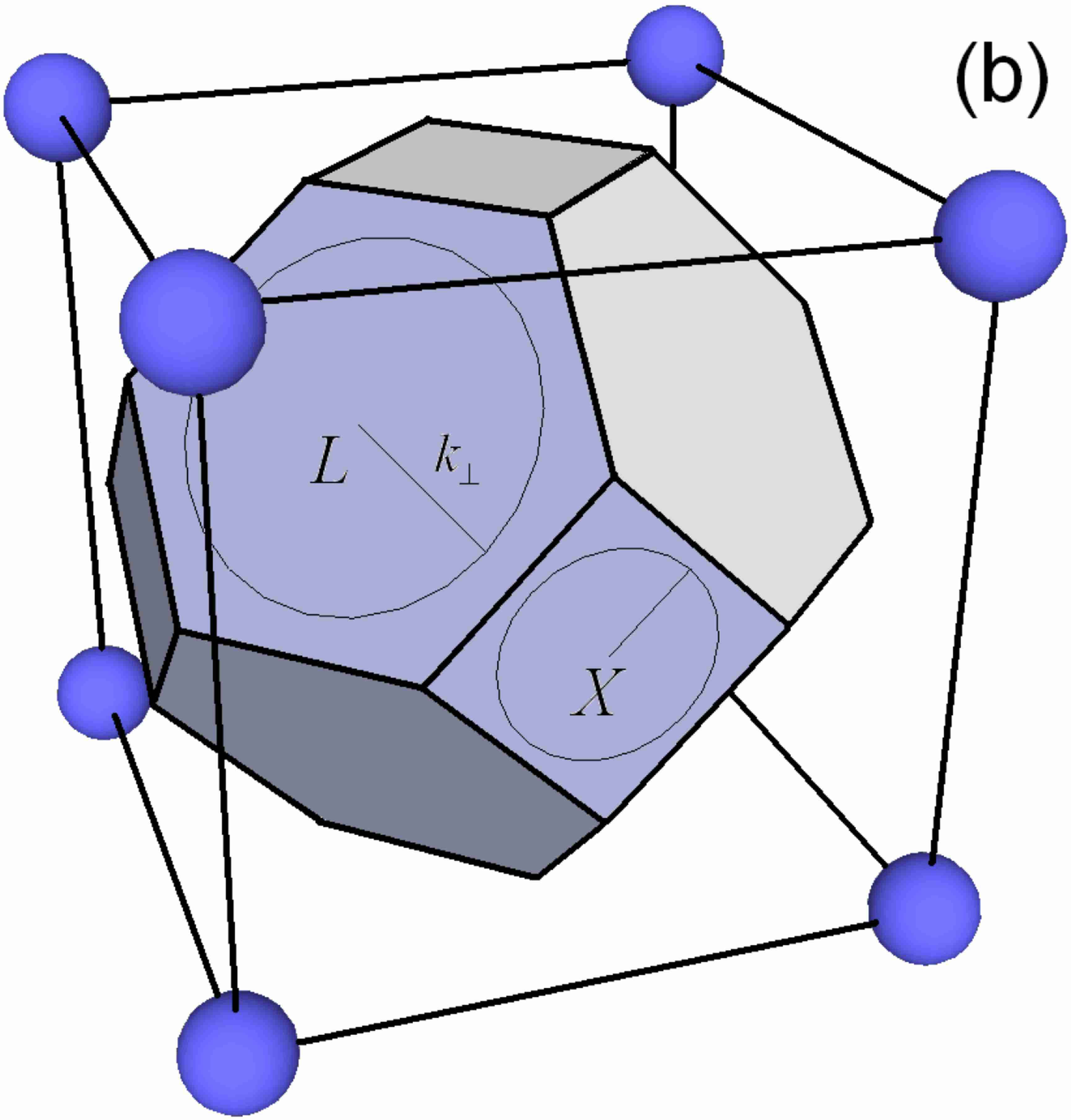}
\includegraphics[width=0.235\textwidth]{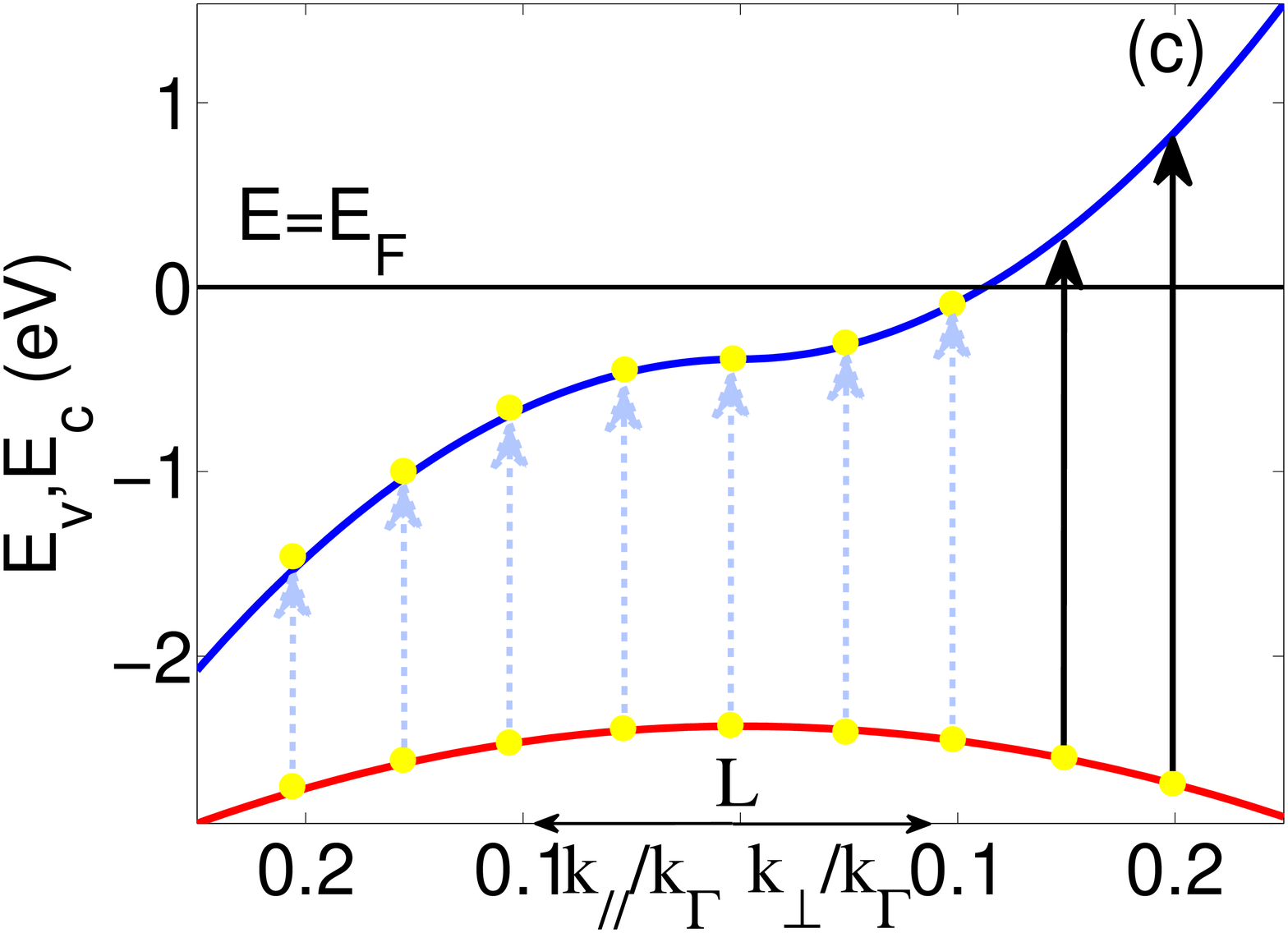}
\includegraphics[width=0.235\textwidth]{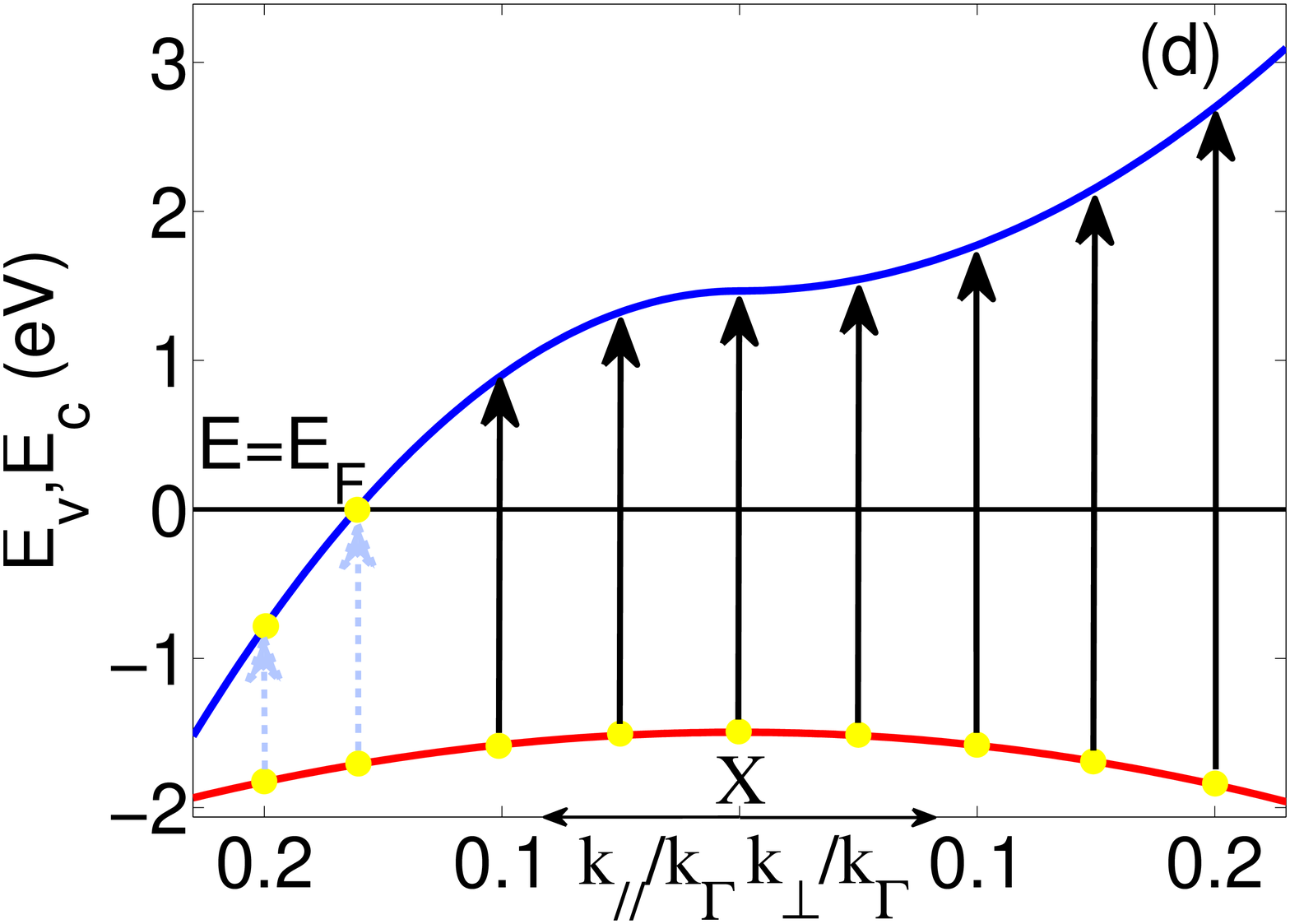}
\caption{(a) Real and (b) reciprocal lattice structures of gold. (c,d) Band structure around the 
(c) $L$ and (d) $X$ points of the reciprocal lattice. The upper blue and lower red curves are the conduction and valence bands, 
respectively. The full black lines represent the Fermi level $E=E_{\rm F}=0$. Note that the right and left sides of the $x$-axis 
correspond to the lattice vector components $k_{\bot},k_{//}$ rescaled to $k_{\Gamma} = 13.34$ nm$^{-1}$.}
\label{Band_Structure_Fig}
\end{center}
\end{figure}

\paragraph{Interband transitions and Bloch equations --} The real and reciprocal lattice structures of gold are face centered 
cubic (f.c.c.) and body centered cubic (b.c.c.), respectively, as shown in Figs. \ref{Band_Structure_Fig}(a,b). Experimental 
results clearly indicate that interband absorption is resonant around the $X,L$ points of the first 
Brillouin zone \cite{ChristensenPRB1971} [see Fig. \ref{Band_Structure_Fig}(b)]. Around these band-edge 
points the Fermi surface is cylindrically simmetric, the lattice vector ${\bf k}$ can be split into in-plane ($k_{\bot}$) 
and out-of-plane ($k_{//}$) components and the valence $E_v({\bf k})$ and conduction $E_c({\bf k})$ bands can be approximated 
by elliptic and hyperbolic paraboloids \cite{MariniNewJPhys2013}. In the following calculations, we approximate the 
conduction band to be completely filled up to the Fermi level $E\le E_{\rm F}$ and completely empty above it $E>E_{\rm F}$. The band 
structure around the $L,X$ points of the reciprocal space is displayed in Figs. \ref{Band_Structure_Fig}(c,d). 
The upper blue and lower red curves are the conduction and valence bands, while the full black lines indicate the Fermi level 
$E=E_{\rm F}=0$. Yellow circles represent occupied states, while full black (dashed cyan) arrows indicate direct interband transitions 
allowed (forbidden) by the Pauli exclusion principle. The modeling of electron interband dynamics is developed as a first order 
perturbative solution of the Schr\"{o}dinger equation in the Heisenberg picture, calculating the time-dependent inversion of
population and the complex polarization from Bloch valence $| v,{\bf k} \rangle$ and conduction $| c,{\bf k} \rangle$ states. 
In our analysis, we neglect the ${\bf k}$-dependence of the dipole matrix element, which is approximated by 
$d_{c,v}({\bf k}) = \langle c, {\bf k} | e {\bf r} | v, {\bf k} \rangle \simeq d_{c,v}$. Thus, the interaction Hamiltonian 
is explicitly given by
\begin{equation}
{\cal H}_{I,{\bf k}} (t) = - E(t)\{ d_{c,v} | c,{\bf k} \rangle \langle v,{\bf k}| + d_{c,v}^* | v,{\bf k} \rangle \langle c,{\bf k}| \},
\end{equation}
where we have used $d_{c,v}^*=d_{v,c}$. From the equation of motion for the density matrix 
$\dot\rho_{\bf k}(t) = -(i/\hbar)[{\cal H}_{\bf k}(t) , \rho_{\bf k}(t)]$, one gets the Bloch equations for the population 
difference $w_{\bf k} = \rho_{c,c}({\bf k},t)-\rho_{v,v}({\bf k},t)$ and for the complex dipole 
$P_{\bf k} = d_{v,c}\rho_{c,v}({\bf k},t)$:
\begin{eqnarray}
\dot w_{\bf k} & = & - \frac{1}{T_1} (1+w_{\bf k}) + \frac{4}{\hbar}E(t)\im{P_{\bf k}}, \label{BlochEq1} \\
\dot P_{\bf k} & = & - \frac{1}{T_2} P_{\bf k} - \mathrm{i} \nu_{\bf k} P_{\bf k} - \frac{\rm{i}}{\hbar} E(t) |d_{v,c}|^2 w_{\bf k}, \label{BlochEq2}
\end{eqnarray}
where $\nu_{\bf k}=[E_c({\bf k})-E_v({\bf k})]/\hbar$ and $T_1 = 500$ fs, $T_2 = 10$ fs are the decay and dephasing 
times of gold. The complex interband polarization ${\bf P}_{II}$ is parallel to the electric 
field ${\bf E}$ and its amplitude is $P_{II}(t) = (2\pi^3)^{-1} \int_{\Omega} P_{\bf k} \mathrm{d}^3 {\bf k}$, where the 
${\bf k}$-integral is taken in the volume $\Omega$ of the reciprocal space where the interband transitions are not forbidden 
by the Pauli exclusion principle. 

\paragraph{Ultrashort SPP pulses in a gold film --} We model optical propagation of SPPs starting from the time-dependent Maxwell 
equations for the electric ${\bf E}$ and magnetic ${\bf H}$ fields. Since the dielectric susceptibility of gold can be separated 
in intraband and interband contributions, it is possible to construct the linear modes with the intraband polarization 
${\bf P}_I$ and study their nonlinear evolution due to the interband counterpart ${\bf P}_{II}$. The space-dependent linear 
susceptibility profile of a thin gold film of width $a$ surrounded by an external dielectric medium is 
$\epsilon_L(x,\omega) = \epsilon_d(\omega) \theta( |x| - a/2 ) + \epsilon_I(\omega) \theta( a/2 - |x| )$,
where $\theta(x)$ is the Heaviside step function, $\epsilon_I(\omega) = 1 + \chi_I(\omega)$ is the intraband 
dielectric constant of gold and $\epsilon_d(\omega)$ is the external dielectric constant.
The space-dependent nonlinear polarization is 
${\bf P}_{\rm NL} = {\bf P}_{II} \theta(a/2-|x|) + (\epsilon_0 \chi_3/2) \left[ |{\bf E}|^2{\bf E} + {\bf E}^2 {\bf E}^*/2 \right] \theta(|x|-a/2)$, 
where $\chi_3$ is the Kerr susceptibility of the external medium 
($\chi_3 = 2.25 \times 10^{-22}$ m$^2$/V$^2$ for silica glass) and 
${\bf P}_{II}$ is the complex interband polarization of gold (implicitly nonlinear) calculated from the 
Bloch equations. In order to derive an effective propagation equation for the field envelope, we proceed 
in a perturbative manner. First, we calculate the linear modes of the plasmonic 
structure neglecting the nonresonant losses ($\epsilon_I''$) and the nonlinear polarization (${\bf P}_{\rm NL}$). 
Then, we perturbatively derive a propagation equation for the linear field envelope accounting for nonresonant 
losses, Kerr nonlinearity and interband polarization. As the structure is assumed to be infinitely extended along the 
$y$-direction, the modal electric ${\bf E}$ and magnetic ${\bf H}$ fields are 
${\bf E} = \sqrt{2 \mu_0 c / S_z }   \psi {\bf e}(x) {\mathrm e}^{ \mathrm{i} \beta_0 z - \mathrm{i} \omega_0 t }$ and
${\bf H} = \sqrt{2 / (\mu_0 c S_z) } \psi {\bf h}(x) {\mathrm e}^{ \mathrm{i} \beta_0 z - \mathrm{i} \omega_0 t }$.
In the {\em Ansatz} above, the fields are normalized to 
$S_z = \int_{-\infty}^{+\infty} {\mathrm Re} [{\bf e}\times {\bf h}^*]\cdot\hat{z} \mathrm{d}x$, $\mu_0,c$ 
are the vacuum magnetic permeability and speed of light, $\omega_0,\beta_0$ are the modal angular frequency and 
wavenumber, ${\bf e}=e_x\hat{x}+e_z\hat{z}$ and 
${\bf h} = (c/\omega_0)(\beta_0e_x+{\mathrm i}{\mathrm d} e_z / {\mathrm d} x )\hat{y}$ 
are the dimensionless electric and magnetic profiles. The modal normalization constants are chosen in such a way that 
$|\psi|^2$ is the power density (measured in W/m) carried in the $z$-direction. 

\begin{figure}
\centering
\begin{center}
\includegraphics[width=0.235\textwidth]{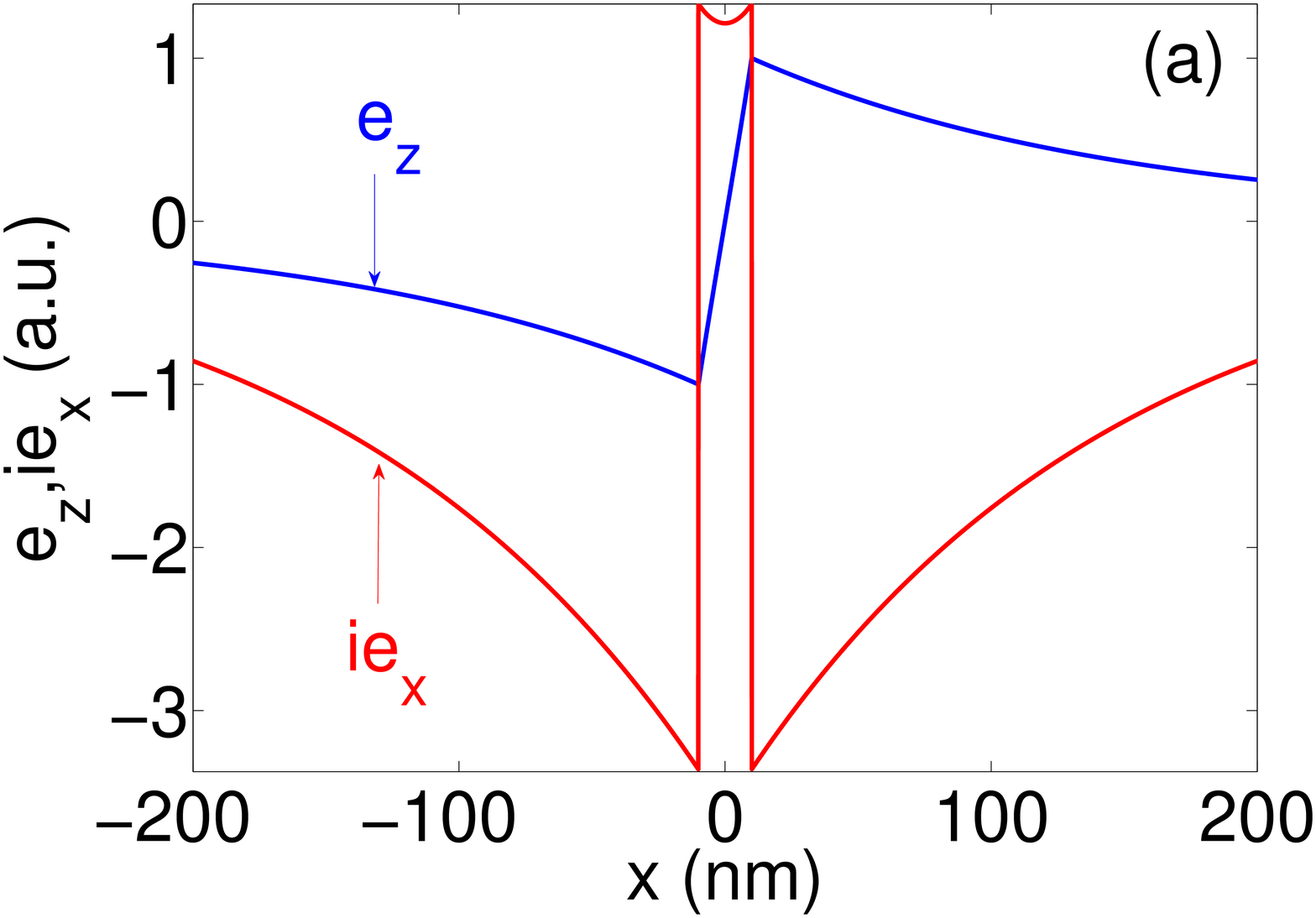}
\includegraphics[width=0.235\textwidth]{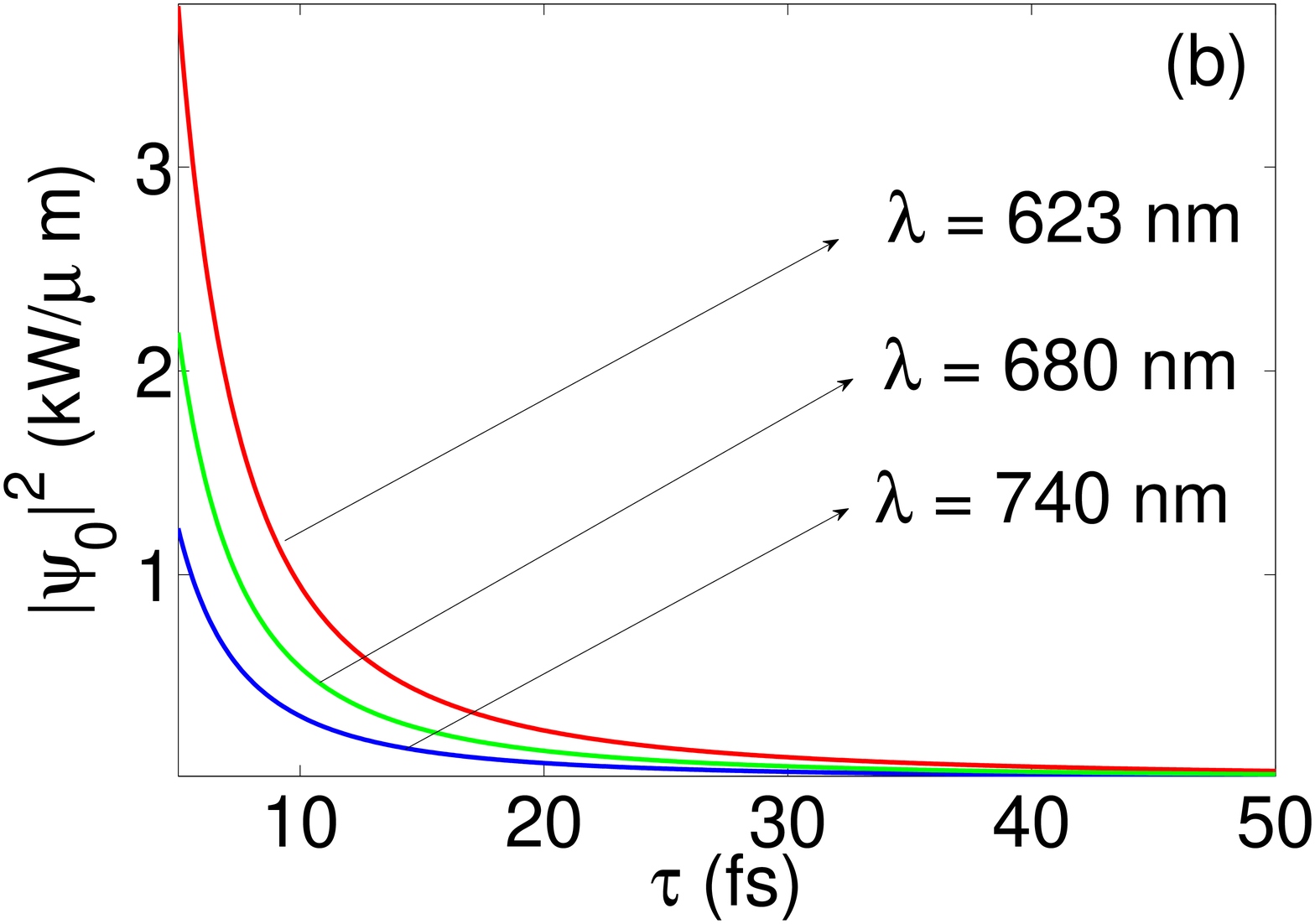}
\caption{(a) Longitudinal ($e_z$) and transverse ($\mathrm{i}e_x$) SPP mode profiles. (b) SIT plasmon soliton peak power density $|\psi_0|^2$ as a function of its time duration $\tau$. In both figures the thickness of the gold film is $a = 20$ nm, the external medium is silica glass and the optical wavelength is $\lambda = 623$ nm.}
\label{psi02_Fig}
\end{center}
\end{figure}

The electric field components $e_x,e_z$ and the propagation constant $\beta_0$ can be calculated by substituting the {\em Ansatz} in the 
time-dependent Maxwell equations and by imposing the boundary conditions (BCs) for the continuity of the longitudinal 
electric field ($e_z$) and transverse displacement vector ($\epsilon'_L e_x$) at the interfaces $x=\pm a/2$.
By doing so, one gets the dispersion relation of LRSPPs ${\rm tanh}( q_m a / 2 ) = - q_d \epsilon'_I / ( q_m \epsilon_d )$,
where $q_{d,m}^2 = \beta_0^2 - \omega^2 \epsilon'_{d,I} / c^2$. The LRSPP longitudinal ($e_z$) and transverse ($\mathrm{i}e_x$)
field profiles are shown in Fig. \ref{psi02_Fig}(a). In the thin film limit $q_m a \ll 1$, 
the transverse electric field profile ($e_x$) is approximately constant and the longitudinal 
electric field profile ($e_z$) is approximatley linear. While the transverse field component 
is never null inside the metal and stays roughly constant, the longitudinal field component vanishes exactly at the center 
of the gold film. For this reason, the major role in the interband dynamics is played by the transverse field component 
that within the gold film can be approximated by $e_x \simeq - 2 \mathrm{i} \beta_0 / ( q_m^2 a )$. In the following step 
of the perturbative analysis, the mode envelope $\psi(z,t)$ is allowed to evolve on a slower scale compared to the fast 
oscillations ${\mathrm e}^{\mathrm{i}\beta_0z-\mathrm{i}\omega_0t}$: $|\partial_z\psi| \ll |\beta_0\psi|$, 
$|\partial_t\psi| \ll |\omega_0\psi|$. Developing a multiscale expansion and applying the Fredholm alternative theorem 
with nonlinearly reinforced BCs \cite{MariniPRA2010} it is possible to prove that one obtains the following generalized nonlinear Schr\"{o}dinger 
Equation (GNLSE) for $\psi(z,t)$:
\begin{eqnarray}
&& \mathrm{i}\partial_z \psi + \mathrm{i}v_g^{-1} \partial_t \psi - (\beta_2/2) \partial_t^2 \psi + \mathrm{i} \alpha \psi  + \nonumber \\
&& + \gamma|\psi|^2\psi + \mathrm{i}\kappa_{\rm p} \int_{\Omega} p_{\bf k} {\mathrm d}^3{\bf k} = 0 , \label{GNLSEQ}
\end{eqnarray}
where $p_{\bf k} = P_{\bf k}{\mathrm e}^{-{\mathrm i}\beta_0z+{\mathrm i}\omega_0t}$, 
$v_g = {\mathrm d} \omega/{\mathrm d} \beta$ is the group velocity, 
$\beta_2 = {\mathrm d}^2 \beta / {\mathrm d}\omega^2 $ 
is the dispersion coefficient, $\alpha$ is the absorption coefficient due to intraband losses,
$\gamma$ is the Kerr nonlinear coefficient and $\kappa_{\rm p}$ is the interband coupling coefficient. The formulation of Eq. (\ref{GNLSEQ}) is the first major result of this Letter.

\begin{figure}
\centering
\begin{center}
\includegraphics[width=0.235\textwidth]{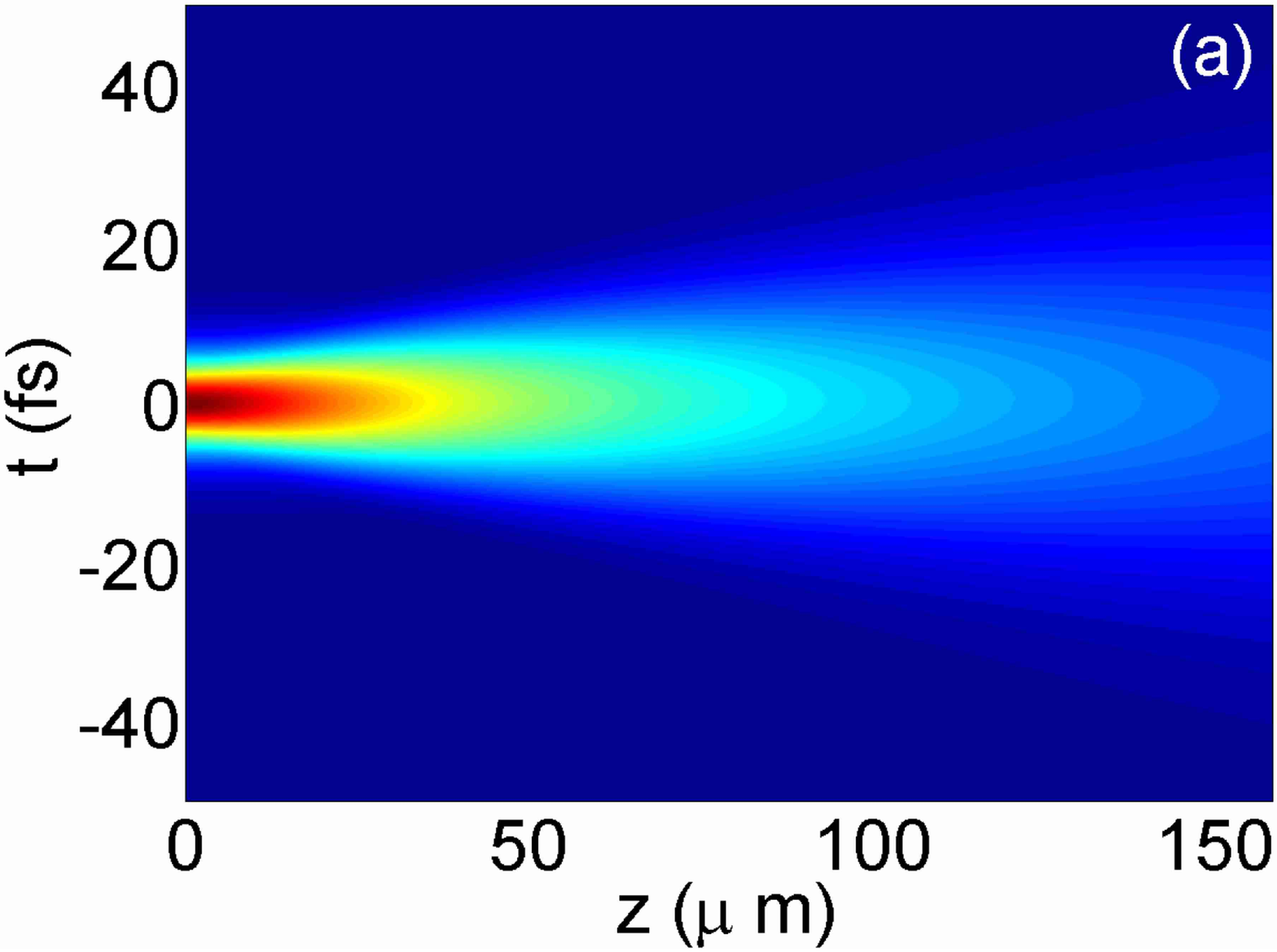}
\includegraphics[width=0.235\textwidth]{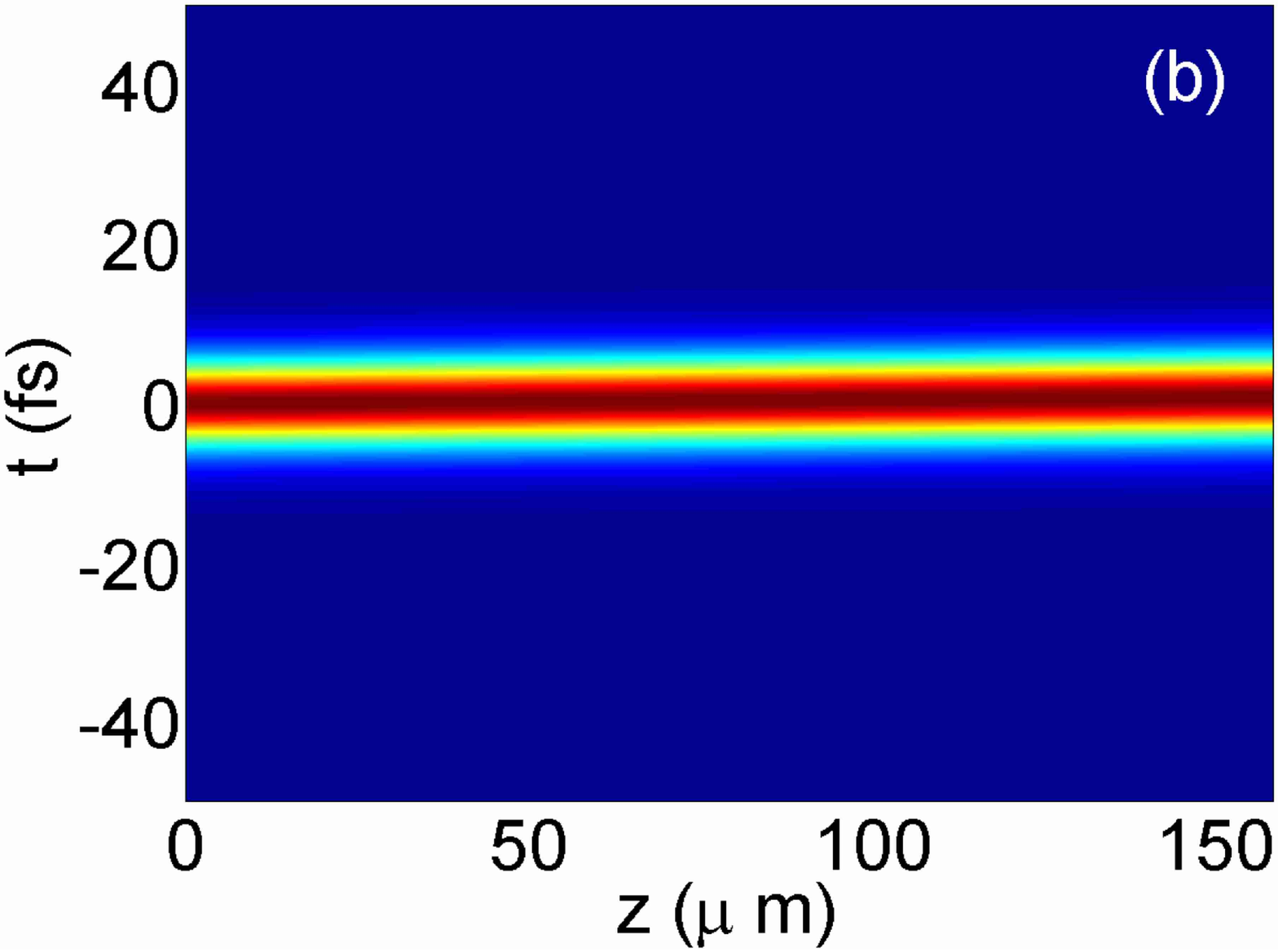}
\includegraphics[width=0.235\textwidth]{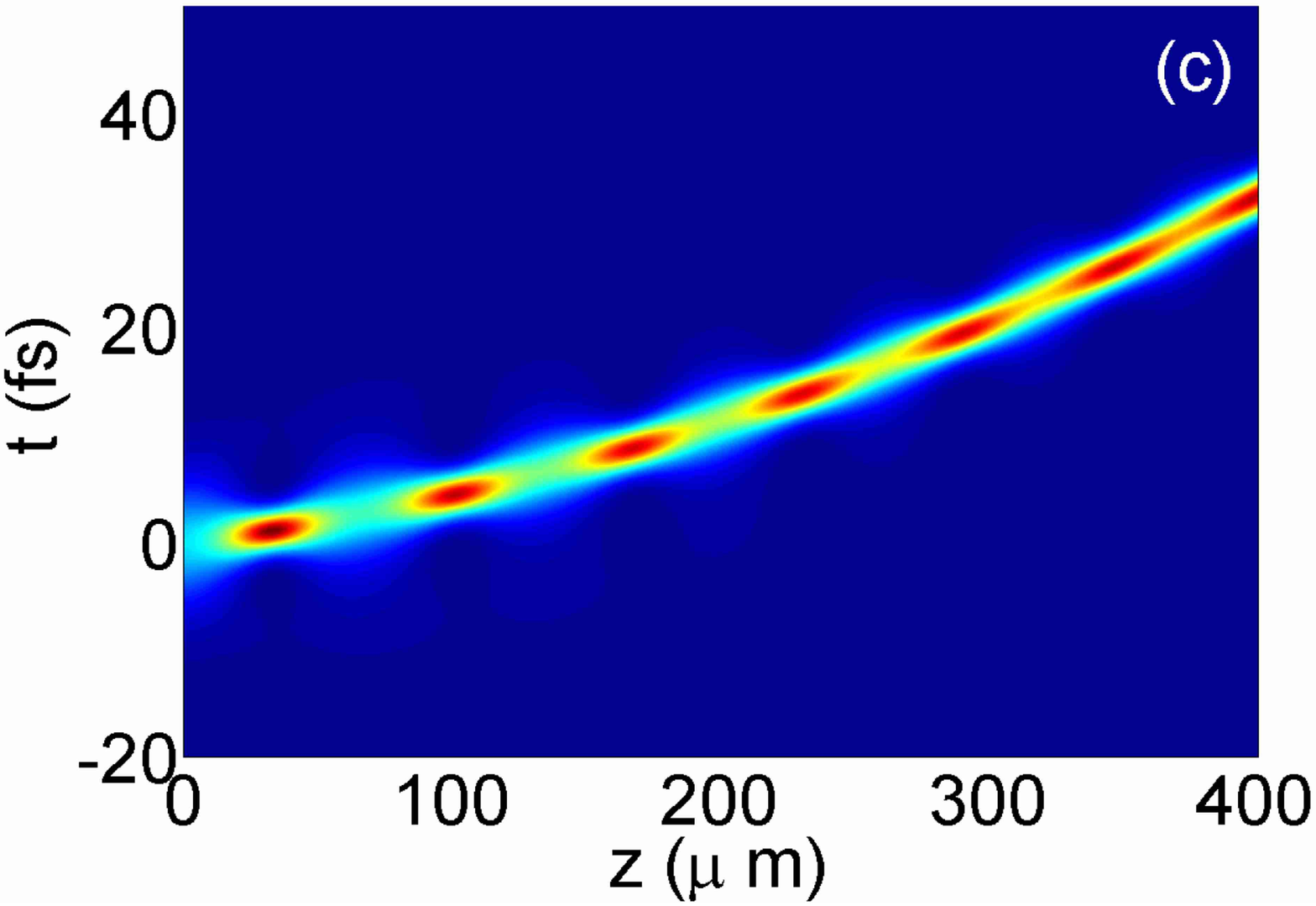}
\includegraphics[width=0.235\textwidth]{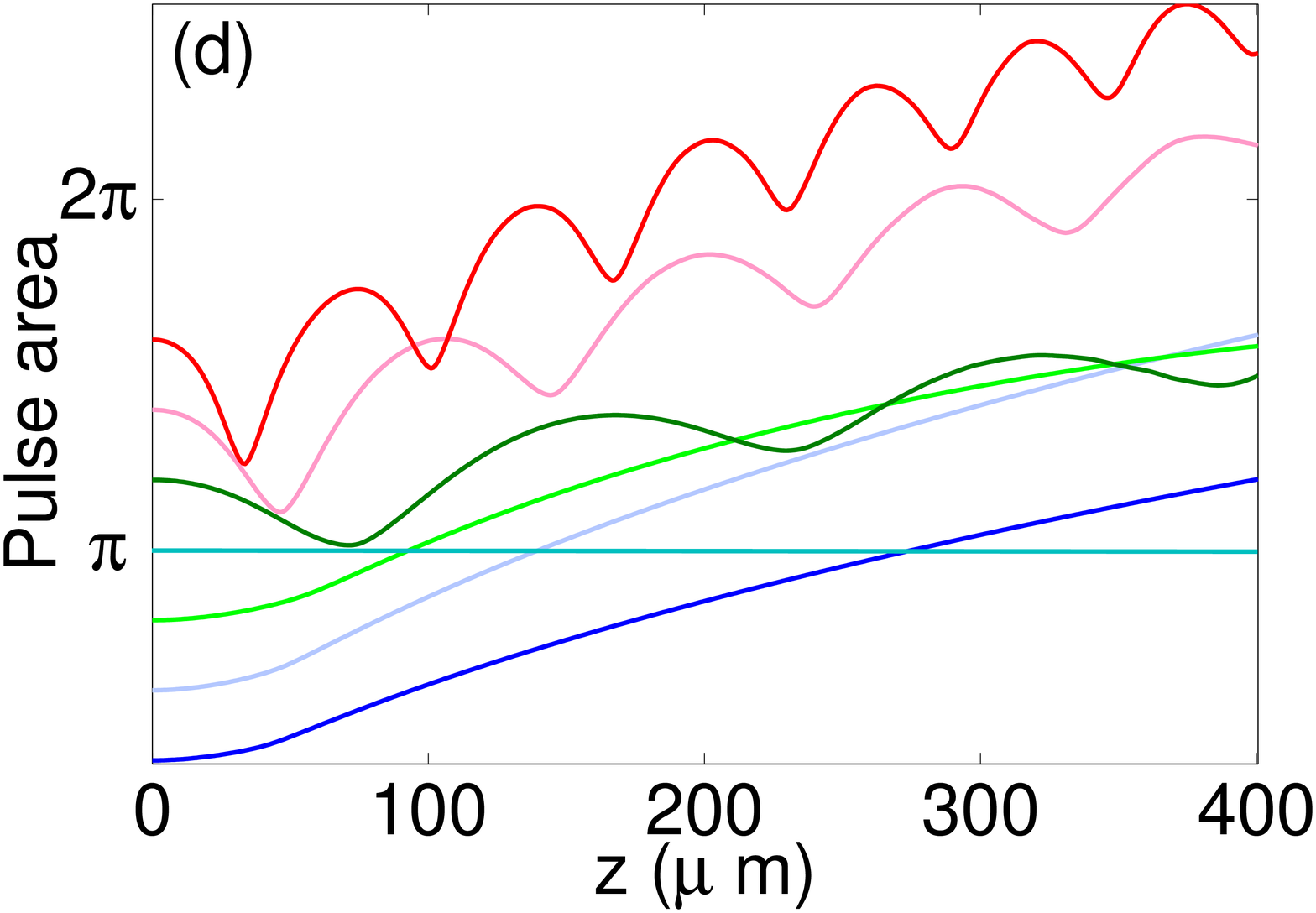}
\caption{Conservative temporal dynamics of hyperbolic secant pulses in absence of intraband loss, dephasing and decay. 
(a,b,c) Temporal evolution of (a,b) $\pi$ and (c) $1.6\pi$ pulses of duration $\tau=5$ fs in a thin film of gold of 
width $a = 20$ nm surrounded 
(a) by silica glass ($\gamma = 3.75 \times 10^{-7}$ W$^{-1}$) and 
(b,c) by a defocusing medium that satisfies the Nakazawa condition 
($\gamma =  - 2.99 \times 10^{-5} W^{-1}$). 
(d) Evolution of pulse area for several input hyperbolic secant pulses with different input amplitude. Note that while the area
of the SIT plasmon soliton remains constantly $\pi$, for other input pulses the pulse area does not remain constant. }
\label{Conservative_Fig}
\end{center}
\end{figure}

\paragraph{Self-induced transmission of SPPs --} The complex interband polarization $p_{\bf k}$ couples the GNLSE with Eqs. (\ref{BlochEq1},\ref{BlochEq2}), which under the rotating wave approximation (RWA) become
\begin{eqnarray}
\dot w_{\bf k} & = & {\cal C} [\psi p_{\bf k}^* + \psi^* p_{\bf k}] - (1+w_{\bf k})/T_1 , \label{BlochRWAEq1} \\
\dot p_{\bf k} & = & {\mathrm i} \Delta_{\bf k} p_{\bf k} - ({\cal C}/2) |d_{v,c}|^2 \psi  w_{\bf k} - p_{\bf k}/T_2 , \label{BlochRWAEq2}
\end{eqnarray}
where $\Delta_{\bf k} = \omega_0 - \nu_{\bf k}$ is the ${\bf k}$-dependent detuning and 
${\cal C} = 2 \beta_0 \sqrt{2 \mu_0 c}/( q_m^2 a\hbar \sqrt{S_z} )$ is the interband coupling constant. 
For pulses with time duration $\tau$ much smaller than decay ($T_1$) and dephasing ($T_2$) times, 
the non-conservative terms of Bloch equations can be disregarded. 
In this limit, neglecting also the intraband loss ($\alpha$), it is possible to derive the expression 
for the family of self-induced transparency (SIT) solitons \cite{McCallPhysRev1969}: 
$\psi(z,t) = \psi_0 {\rm sech}\left[ (t-z/{\cal V})/ \tau \right]{\mathrm e}^{\mathrm{i}\Delta\beta z}$,
where $\psi_0 = 2 \left( {\cal C} |d_{v,c}| \tau \right)^{-1}$, ${\cal V}$ is the SIT soliton velocity and 
$\Delta\beta$ is its phase shift. The dependence of the SIT plasmon soliton power density ($|\psi_0|^2$) on 
its time duration ($\tau$) is depicted in Fig. \ref{psi02_Fig}(b).
Assuming that the lateral confinement in the unbound $y$-direction is of the order of $L_y\approx 10$ $\mu$m,
one gets that SIT plasmon solitons can be excited with peak powers of the order of $P=|\psi_0|^2L_y\simeq 10$ kW.
We emphasize that the SIT plasmon soliton exists only for a particular coincidence of geometrical parameters such
that the GVD is exactly compensated for by the Kerr nonlinearity $\beta_2 {\cal C}^2 |d_{v,c}|^2 + 4 \gamma = 0$. 
The interband absorption is resonant at $\lambda = 400$ nm and has a characteristic spectral width of $\Delta\lambda\simeq 100$ nm \cite{MariniNewJPhys2013}. Thus, for $\lambda \gtrsim 500$ nm the band dispersion does not play a crucial role, it is possible 
to approximate $\Delta_{\bf k}\simeq \Delta$ and the interband dynamics is simplified to a two-level system with effective 
detuning $\Delta$. In this approximation, we have solved numerically 
Eqs. (\ref{GNLSEQ},\ref{BlochRWAEq1},\ref{BlochRWAEq2}) by using the split-step Fourier method with fourth-order Runge-Kutta algorithms for the spatial propagation and the Adams-Bashforth method for the temporal integration of Bloch equations. 
In the simulations, we considered a thin film of gold with thickness $a=20$ nm surrounded by silica glass at the optical 
wavelength $\lambda = 623$ nm, characterized by group velocity $v_g = 194$ km/s and dispersion coefficient $\beta_2 = 0.92$ $\mu$m$^{-1}$ fs$^2$. 
Results of numerical simulations in the conservative limit are reported in the panel of Fig. \ref{Conservative_Fig}. 
In Figs. \ref{Conservative_Fig}(a,b,c) we show the temporal dynamics of hyperbolic secant pulses of duration $\tau=5$ fs 
with area (a,b) $\pi$ and (c) $1.6\pi$ in a thin film of gold of width $a = 20$ nm surrounded 
(a) by silica glass ($\gamma = 3.75 \times 10^{-7}$ W$^{-1}$) and 
(b,c) by a defocusing medium that satisfies the soliton condition existence $\beta_2 {\cal C}^2 |d_{v,c}|^2 + 4 \gamma = 0$, with
$\gamma =  - 2.99 \times 10^{-5}$ W$^{-1}$.
In Fig. \ref{Conservative_Fig}(d), the evolution of the pulse area is depicted for several input hyperbolic secant pulses with
different amplitude. Note that, when the soliton existence condition is satisfied, the propagation of $\pi$ pulses does not affect 
their shape, while pulses with different areas are either absorbed or undergo an oscillatory dynamics. The discovery of such ultrashort SIT solitons in gold is the second major result of this Letter.

\begin{figure}
\centering
\begin{center}
\includegraphics[width=0.235\textwidth]{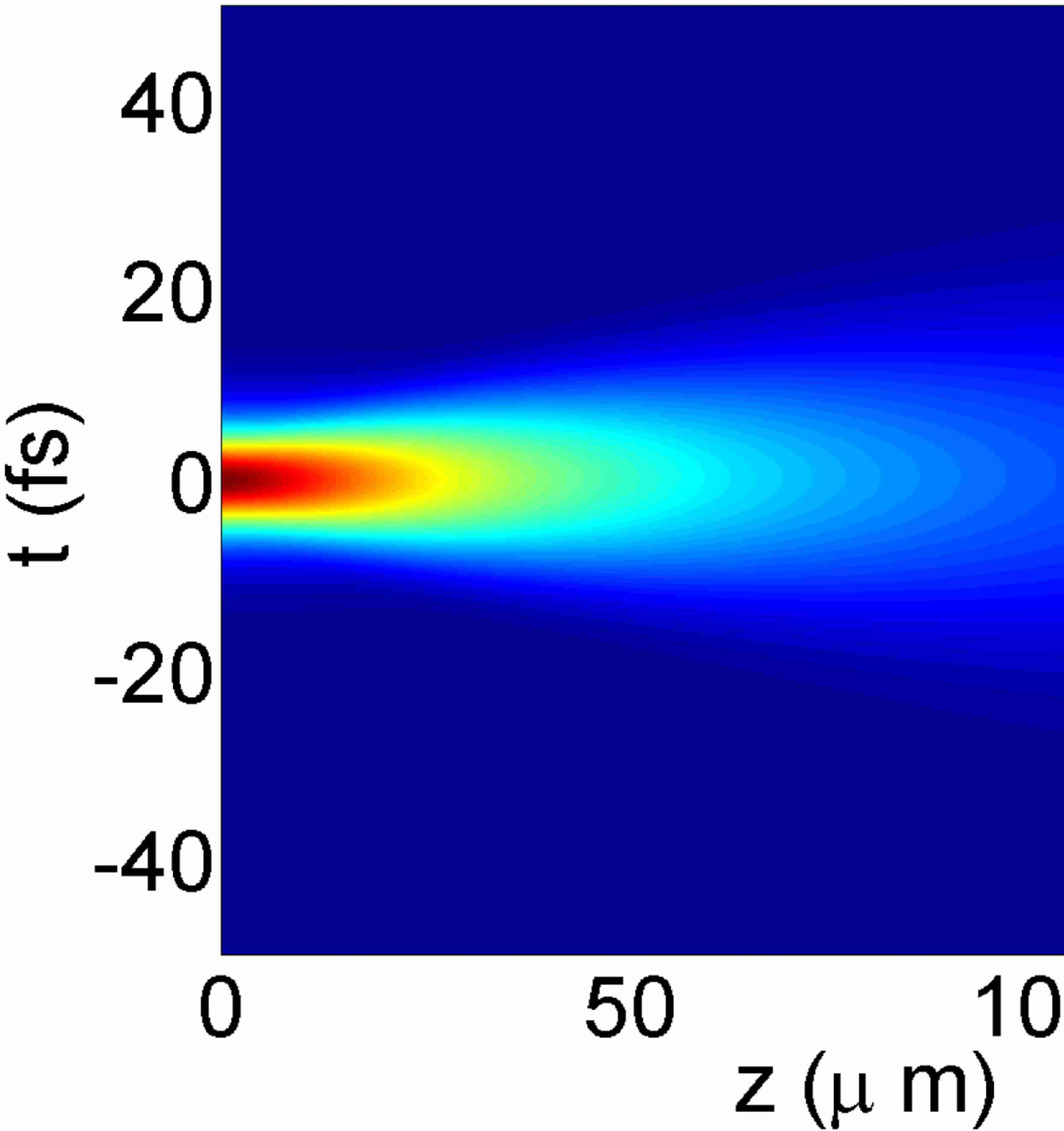}
\includegraphics[width=0.235\textwidth]{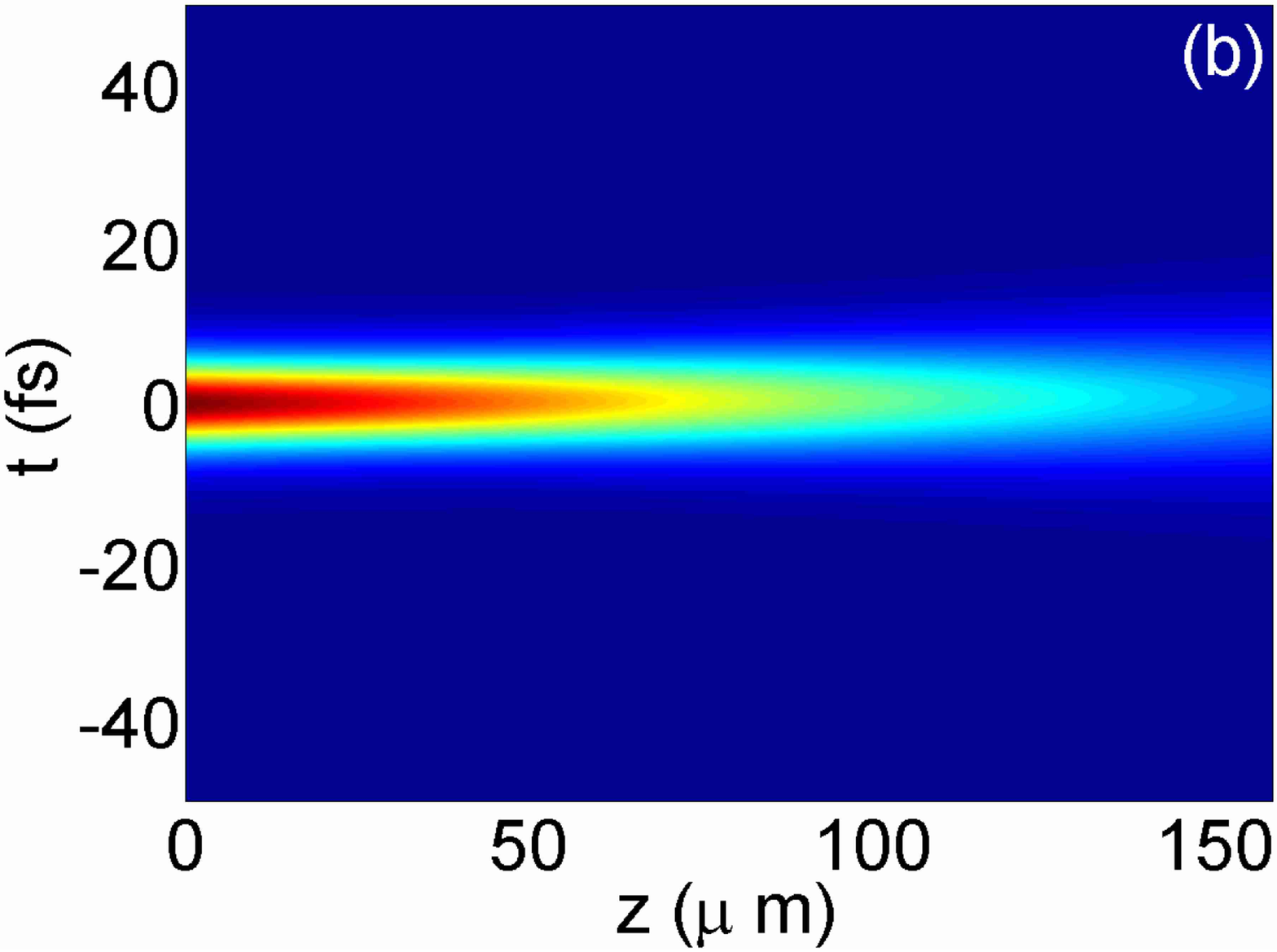}
\includegraphics[width=0.235\textwidth]{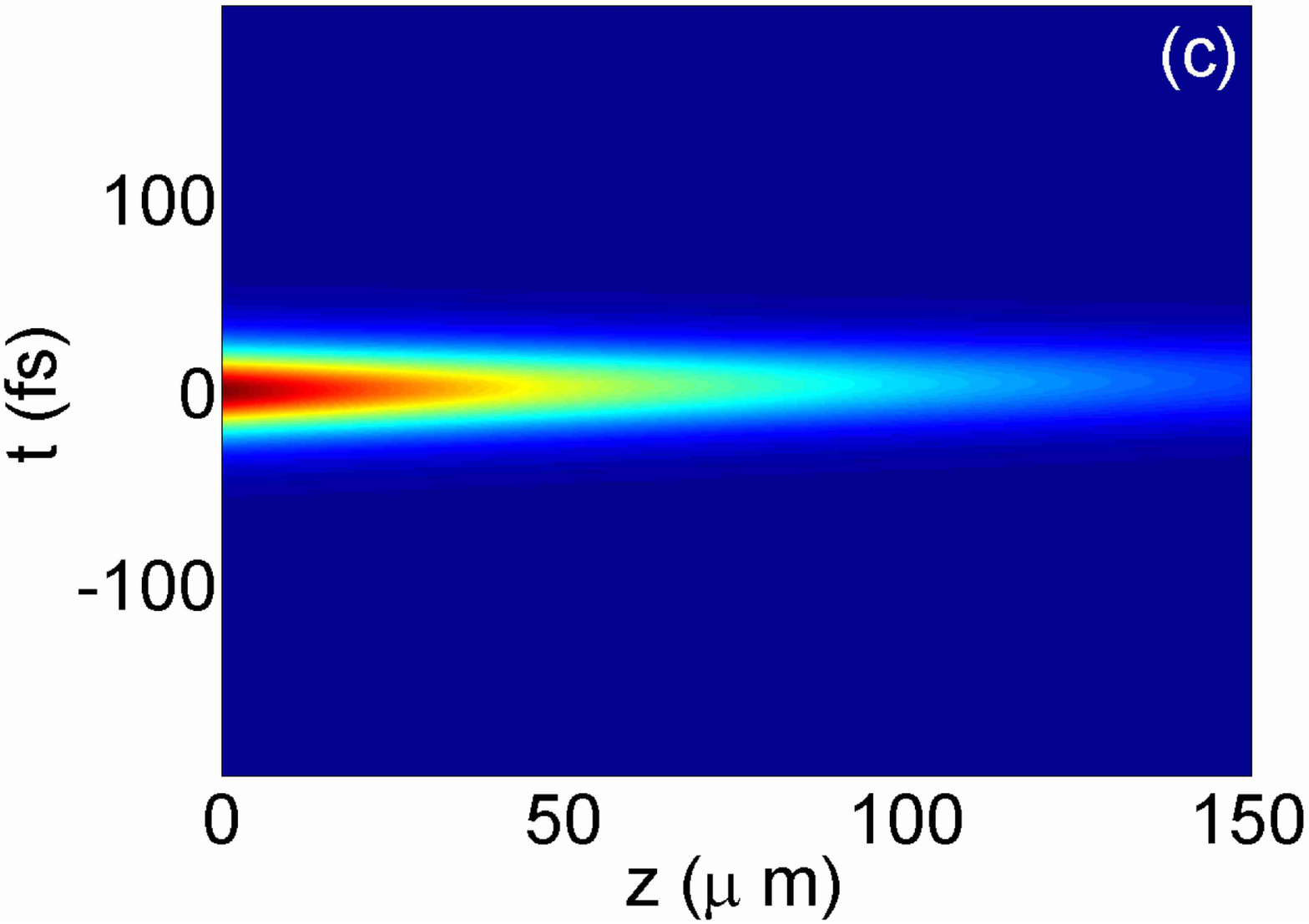}
\includegraphics[width=0.235\textwidth]{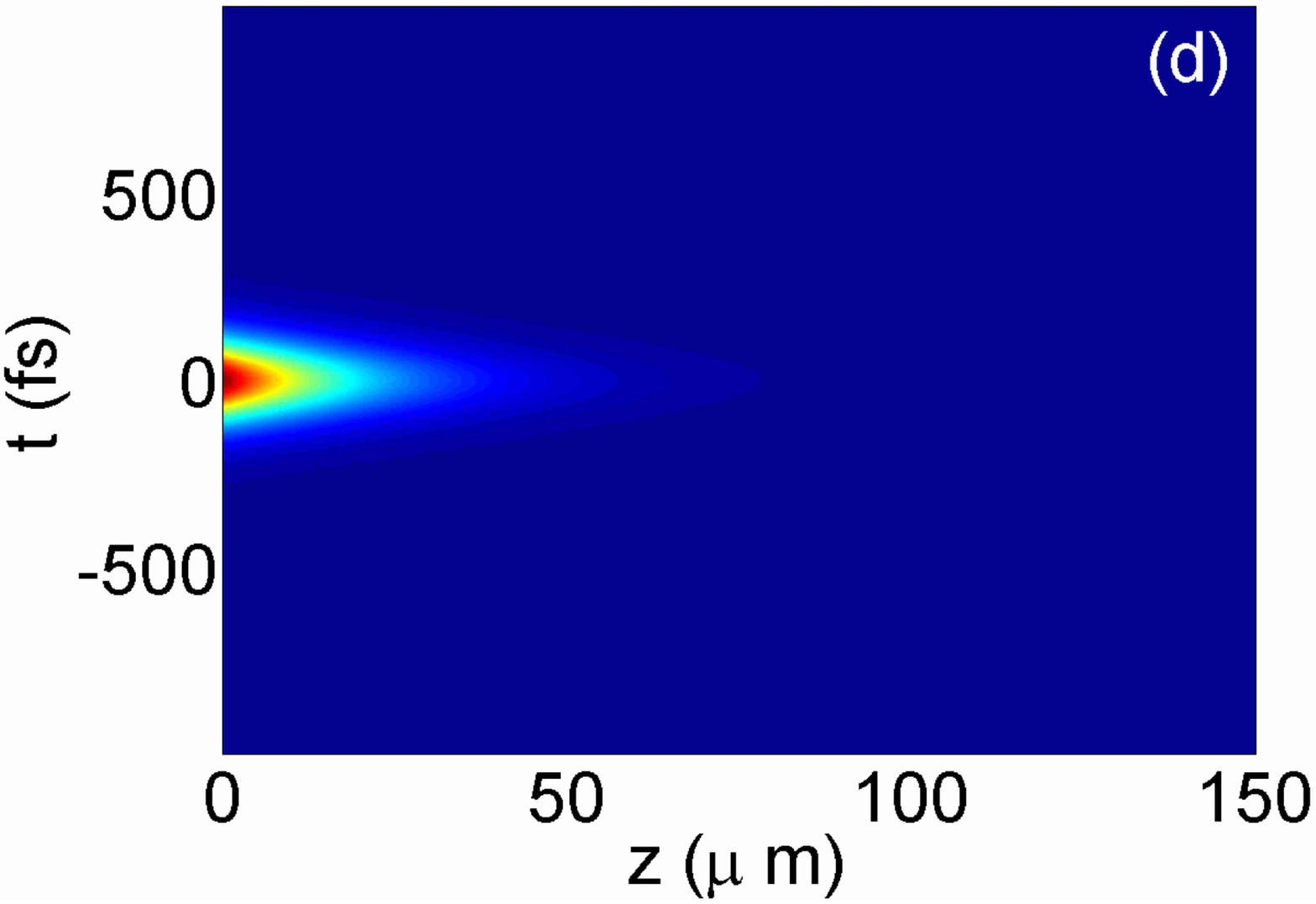}
\caption{Non-conservative temporal dynamics of hyperbolic secant pulses. Temporal evolution of $\pi$ pulses of durations 
(a,b) $\tau = 5$ fs, (c) $\tau = 20$ fs and (d) $\tau = 100$ fs in a thin film of gold of width $a = 20$ nm surrounded 
(a) by silica glass ($\gamma = 3.75 \times 10^{-7}$ W$^{-1}$) and 
(b,c,d) by a defocusing medium that satisfies the soliton existence condition
($\gamma =  - 2.99 \times 10^{-5}$ W$^{-1}$).
}
\label{NonConservative_Fig}
\end{center}
\end{figure}

Results of numerical simulations in realistic non-conservative conditions are reported in the panel of Fig. \ref{Conservative_Fig}. 
In Figs. \ref{Conservative_Fig}(a,b,c) we show the temporal dynamics of hyperbolic secant pulses of duration $\tau=5$ fs 
with $\pi$ area in a thin film of gold surrounded of width $a = 20$ nm
(a) by silica glass ($\gamma = 3.75 \times 10^{-7} W^{-1}$) and 
(b,c) by a defocusing medium that satisfies the soliton condition existence $\beta_2 {\cal C}^2 |d_{v,c}|^2 + 4 \gamma = 0$: 
$\gamma =  - 2.99 \times 10^{-5}$ W$^{-1}$.
In Fig. \ref{Conservative_Fig}(d), the evolution of a long pulse of duration $\tau = 100$ fs is depicted. 
In this limit thermalization occurs and the optical propagation is modeled in terms of an effective dielectric constant. 
Note that, even if ideal transparency is not achieved, ultrashort pulses of duration $\tau \simeq 10$ fs propagate over
much longer distance with respect to long pulses of duration $\tau \simeq 100$ fs. 

\paragraph{Conclusions --} We have investigated analytically and numerically
the interband self-induced transmission of surface plasmon polaritons in a gold film, demonstrating
the possibility to bypass the interband losses of gold by using ultrashort $\pi$-pulses.
This regime is realistically accessible to experiments with pulse peak powers of the order
of $P\simeq 10$ kW and pulse time duration $\tau \le 10$ fs. These theoretical findings highlight
the novel possibilities offered by the use of ultrashort pulses in plasmonic systems and can
stimulate further developments and strategies to reduce the impact of losses in metals, as well as opening up new directions for the exploitation of nonlinear effects in metals.


\begin{thebibliography}{0}
\expandafter\ifx\csname natexlab\endcsname\relax\def\natexlab#1{#1}\fi
\expandafter\ifx\csname bibnamefont\endcsname\relax
  \def\bibnamefont#1{#1}\fi
\expandafter\ifx\csname bibfnamefont\endcsname\relax
  \def\bibfnamefont#1{#1}\fi
\expandafter\ifx\csname citenamefont\endcsname\relax
  \def\citenamefont#1{#1}\fi
\expandafter\ifx\csname url\endcsname\relax
  \def\url#1{\texttt{#1}}\fi
\expandafter\ifx\csname urlprefix\endcsname\relax\def\urlprefix{URL }\fi
\providecommand{\bibinfo}[2]{#2}
\providecommand{\eprint}[2][]{\url{#2}}

\end{thebibliography}


\begin{thebibliography}{10}
\newcommand{\enquote}[1]{``#1''}

\bibitem{GobinNanoLett2007} A. M. Gobin, M. H. Lee, N. J. Halas, W. D. James, R. A. Drezek and J. L. West, Nano Lett. {\bf 7}, 1929 (2007).

\bibitem{AnkerNatMat2008} J. N. Anker, W. P. Hall, O. Lyandres, N. C. Shah, J. Zhao and R. P. Van Duyne, Nat. Mat. {\bf 7}, 442 (2008).

\bibitem{KawataNatPhot2009} S. Kawata, Y. Inouye and P. Verma, Nat. Phot. {\bf 3}, 388 (2009).

\bibitem{GramotnevNatPhot2010} D. K. Gramotnev and S. I. Bozhevolnyi, Nat. Phot. {\bf 4}, 83 (2010).

\bibitem{GinzburgNewJPhys2013} P. Ginzburg, A. V. Krasavin and A. V. Zayats, New J. Phys. {\bf 15}, 013031 (2013).

\bibitem{DavoyanOL2011} A. R. Davoyan, I. V. Shadrivov and Y. S. Kivshar, Opt. Lett. {\bf 36}, 930 (2011).

\bibitem{MariniNewJPhys2013} A. Marini, M. Conforti, G. Della Valle, H. W. Lee, Tr. X. Tran, W. Chang, M. A. Schmidt, S. Longhi, P. St.J. Russell and F. Biancalana, New J. Phys. {\bf 15}, 013033 (2013).

\bibitem{KauranenNatPhot2012} M. Kauranen and A. V. Zayats, Nat. Phot. {\bf 6}, 737 (2012).

\bibitem{Feigenbaum2007} E. Feigenbaum and M. Orenstein, Opt. Lett. {\bf 32}, 674 (2007).

\bibitem{DavoyanPRL2010} A. R. Davoyan, I. V. Shadrivov, A. A. Zharov, D. K. Gramotnev and Y. S. Kivshar, Phys. Rev. Lett. {\bf 105}, 116804 (2010).

\bibitem{GatherNature2010} M. C. Gather, K. Meerholz, N. Danz and K. Leosson, Nat. Phot. {\bf 4}, 457 (2010).

\bibitem{MariniPRA2010} A. Marini and D. V. Skryabin, Phys. Rev. A {\bf 81}, 033850 (2010).

\bibitem{BeriniAdvOpt2009} P. Berini, Adv. in Opt. and Phot. {\bf 1}, 484 (2009).

\bibitem{McCallPhysRev1969} S. L. McCall and E. L. Hahn, Phys. Rev. {\bf 183}, 457 (1969).

\bibitem{NakazawaPRL1991} N. Nakazawa, E. Yamada and H. Kubota, Phys. Rev. Lett. {\bf 66}, 2625 (1991).

\bibitem{ChristensenPRB1971} N. E. Christensen and B. O. Seraphin, Phys. Rev. B {\bf 4}, 3321 (1971).

\end{thebibliography}
\end{document}